\input harvmac.tex

\def\unlockat{\catcode`\@=11}
\def\lockat{\catcode`\@=12}
\unlockat
\def\newsec#1{\global\advance\secno by1\message{(\the\secno. #1)}
\global\subsecno=0\global\subsubsecno=0
\global\deno=0\global\prono=0\global\teno=0\eqnres@t\noindent
{\bf\the\secno. #1} \writetoca{{\secsym}
{#1}}\par\nobreak\medskip\nobreak}
\global\newcount\subsecno \global\subsecno=0
\def\subsec#1{\global\advance\subsecno
by1\message{(\secsym\the\subsecno. #1)}
\ifnum\lastpenalty>9000\else\bigbreak\fi\global\subsubsecno=0
\global\deno=0\global\prono=0\global\teno=0
\noindent{\it\secsym\the\subsecno. #1} \writetoca{\string\quad
{\secsym\the\subsecno.} {#1}}
\par\nobreak\medskip\nobreak}
\global\newcount\subsubsecno \global\subsubsecno=0
\def\subsubsec#1{\global\advance\subsubsecno by1
\message{(\secsym\the\subsecno.\the\subsubsecno. #1)}
\ifnum\lastpenalty>9000\else\bigbreak\fi
\noindent\quad{\secsym\the\subsecno.\the\subsubsecno.}{#1}
\writetoca{\string\qquad{\secsym\the\subsecno.\the\subsubsecno.}{#1}}
\par\nobreak\medskip\nobreak}
\global\newcount\deno \global\deno=0
\def\de#1{\global\advance\deno by1
\message{(\bf Definition\quad\secsym\the\subsecno.\the\deno #1)}
\ifnum\lastpenalty>9000\else\bigbreak\fi \noindent{\bf
Definition\quad\secsym\the\subsecno.\the\deno}{#1}
\writetoca{\string\qquad{\secsym\the\subsecno.\the\deno}{#1}}}
\global\newcount\prono \global\prono=0
\def\pro#1{\global\advance\prono by1
\message{(\bf Proposition\quad\secsym\the\subsecno.\the\prono #1)}
\ifnum\lastpenalty>9000\else\bigbreak\fi \noindent{\bf
Proposition\quad\secsym\the\subsecno.\the\prono}{#1}
\writetoca{\string\qquad{\secsym\the\subsecno.\the\prono}{#1}}}
\global\newcount\teno \global\prono=0
\def\te#1{\global\advance\teno by1
\message{(\bf Theorem\quad\secsym\the\subsecno.\the\teno #1)}
\ifnum\lastpenalty>9000\else\bigbreak\fi \noindent{\bf
Theorem\quad\secsym\the\subsecno.\the\teno}{#1}
\writetoca{\string\qquad{\secsym\the\subsecno.\the\teno}{#1}}}
\def\subsubseclab#1{\DefWarn#1\xdef
#1{\noexpand\hyperref{}{subsubsection}%
{\secsym\the\subsecno.\the\subsubsecno}%
{\secsym\the\subsecno.\the\subsubsecno}}%
\writedef{#1\leftbracket#1}\wrlabeL{#1=#1}}
\lockat
\def\IB{\relax\hbox{$\inbar\kern-.3em{\rm B}$}}
\def\IC{\relax\hbox{$\inbar\kern-.3em{\rm C}$}}
\def\ID{\relax\hbox{$\inbar\kern-.3em{\rm D}$}}
\def\IE{\relax\hbox{$\inbar\kern-.3em{\rm E}$}}
\def\IF{\relax\hbox{$\inbar\kern-.3em{\rm F}$}}
\def\IG{\relax\hbox{$\inbar\kern-.3em{\rm G}$}}
\def\IGa{\relax\hbox{${\rm I}\kern-.18em\Gamma$}}
\def\IH{\relax{\rm I\kern-.18em H}}
\def\IK{\relax{\rm I\kern-.18em K}}
\def\IL{\relax{\rm I\kern-.18em L}}
\def\IP{\relax{\rm I\kern-.18em P}}
\def\IR{\relax{\rm I\kern-.18em R}}
\def\IZ{\relax\ifmmode\mathchoice
{\hbox{\cmss Z\kern-.4em Z}}{\hbox{\cmss Z\kern-.4em Z}}
{\lower.9pt\hbox{\cmsss Z\kern-.4em Z}} {\lower1.2pt\hbox{\cmsss
Z\kern-.4em Z}}\else{\cmss Z\kern-.4em Z}\fi}
\def\II{\relax{\rm I\kern-.18em I}}

\def\frac#1#2{{#1\over#2}}
\def\pr {\partial}
\def\apr {\overline {\partial }}

\def\refb[1] {{(\ref{\#1})}}
\def\eq[1] {{eq.(\ref{\#1}) }}
\def\bear {{\begin{array}}}

\def\qq[1] {{\frac{\overline{\pr ^2 {\cal F}}}{\pr z^{\overline {\#1}}}}}
\def\R {{\bf R}}
\def\C {{\bf C}}

\def\Z {{\bf Z}}

\def\inv[1] {{{\#1}^{-1}}} 

\def\F{\pr{\cal F}(z)}
\def\FF{\pr^2{\cal F}}
\def\FFF{\pr^3{\cal F}}
\def\GG{g}

\def\FFFs{\pr^3{{\cal F}(z_*)}}

\def\aGG{g}
\def\aGGs{g(z_*)}

\def\NN {{\cal N}}

\def\AAA {{\overline{A}}}
\def\ai {{\overline{i}}}
\def\aj {{\overline{j}}}
\def\ak {{\overline{k}}}
\def\az {{\overline{z}}}
\def\z {{\overline{z}}}
\def\W {{\mathcal{W}}}

\def\MM {{\cal M}}

\def\pr {{\partial}}
\def\apr {{\bar \partial}}
\def\az {{\bar z}}
\def\ai {{\bar i}}

\def\g {{\gamma}}

\def\l {{\lambda}}




\def\inbar{\,\vrule height1.5ex width.4pt depth0pt}
\font\cmss=cmss10 \font\cmsss=cmss10 at 7pt

\font\manual=manfnt \def\dbend{\lower3.5pt\hbox{\manual\char127}}

\def\boxit#1{\vbox{\hrule\hbox{\vrule\kern8pt
\vbox{\hbox{\kern8pt}\hbox{\vbox{#1}}\hbox{\kern8pt}}
\kern8pt\vrule}\hrule}}
\def\mathboxit#1{\vbox{\hrule\hbox{\vrule\kern8pt\vbox{\kern8pt
\hbox{$\displaystyle #1$}\kern8pt}\kern8pt\vrule}\hrule}}
\lref\BCOV{M. Bershadsky, S. Cecotti, H. Ooguri and  C. Vafa,
``{\it Kodaira-Spencer Theory of Gravity and Exact Results
for Quantum String Amplitudes},'' Commun.Math.Phys. 165 (1994) 311-428.}
\lref\W{E. Witten, ``{\it Quantum Background Independence In String
    Theory},''
  hep-th/9306122.}
\lref\Hit{N. Hitchin, ``{\it The geometry of three-forms in six and seven
    dimensions},''
\vskip 0.01cm
  math.DG/0010054.}
\lref\Hitone{N. Hitchin, ``{\it Stable forms and special metrics},''
 math.DG/0107101.}
\lref\Hittwo{N. Hitchin, ``{\it The moduli space of complex
Lagrangian submanifolds},''
\vskip 0.01cm
math.DG/9701069. }
\lref\Vtwo{M. Aganagic, R. Dijkgraaf, A. Klemm, M.Marino and C. Vafa,
`` {\it Toplogical Strings and Integrable Hierarchies},'' hep-th/0312085.}
\lref\V{M. Bershadsky, S. Cecotti, H. Ooguri and C. Vafa, ``{\it
Holomorphic anomaly in Topological Field Theories},''
 Nucl.Phys. {\bf B405} (1993) 279, hep-th/9302103.}
\lref\SS{ S. Shatashvili, talks at: Ludwigfest, Leonard
Euler Institute, St. Petersburg, 1994;
IHES and CERN 1995; Ludwigfest, Erwin Schrodinger Institute,
Vienna, March 22, 2004,
http://www.esi.ac.at/activities/archive/Ludwigfest2004.html.
}
\lref\Hthree{N. Hitchin, ``{\it Generalized Calabi-Yau manifolds},''
 math.DG/0209099.}
\lref\dWvP{B. de Witt and  A. Van Proeyen,
``{\it Potentials and Symmetries of general gauged $N=2$ supergravity-Yang-Mills
models}, ``Nucl. Phys.  {\bf B245}  (1984), 89.}
\lref\F{ D. Freed, ``{\it Special Kahler Manifolds},''
  Commun.Math.Phys. 203 (1999) 31.}
\lref\Kon{ M. Kontsevich, `` {\it Intersection Theory on the Moduli
    Space of Curves and the Matrix Airy Function},''
    Commun. Math. Phys. 147 (1992) 1.}
\lref\AG{A. Gerasimov,  talk at "Frobenius manifolds, Quantum
cohomology and singularity", MPIM, July, 2002.}
\lref\GS{A. Gerasimov and  S. Shatashvili, ``{\it Generalized anomaly
equation in
Frobenius geometry}''.}
\lref\AT{A. Todorov, ``{\it Witten's Geometric Quantization on the Moduli
    of CY Threefolds},''  math.AG/0004043.}
\lref\DVV{R. Dijkgraaf, E. Verlinde and  M. Vonk, ``{\it On the partition
sum of
  the NS five-brane},''  hep-th/0205281.}
\lref\GStwo{A. Gerasimov and  S. Shatashvili, ``{\it Towards Integrability
of
Topological Strings II:
AdS/CFT patterns in topological strings and exact solution},'' to
appear. }
\lref\NOV{N. Nekrasov, H. Ooguri and  C. Vafa,  ``{\it S-duality and
Topological
Strings},''
\vskip 0.01cm
hep-th/0403167.}
\lref\DM{D. Morrison, ``{\it Mirror symmetry and rational curves on
    quintic threefolds: a guide for mathematicians},''
  J. Amer. Math. Soc. 6 (1993) 223, alg-geom/9202004.}
\lref\HV{H. Verlinde, ``{\it Conformal Filed Theory, 2-d Quantum Gravity
 and Quantization of Teichmuller Space},''  Nucl.Phys. {\bf B337}
(1990)  652.}
\lref\Wtwo{E. Witten, `` {\it (2+1)-Dimensional Gravity as an Exactly
    Soluble system},''
\vskip 0.01cm
 Nucl.Phys. {\bf B311} (1988) 46. }
\lref\BR{M.J. Bowick and  S.G. Rajeev, ``{\it String theory
as the K\"{a}hler geometry of loop space},''  Phys.Rev.Lett 58 (1987)
535.}
\lref\DT{S. Donaldson and  R. Thomas, ``{\it Gauge theory in higher
    dimensions},''in ``The geometric universe:science, geometry and the
work
of Roger Penrose'' (S.A. Hugget et al, eds.) Oxford Univ. press, 1998,
31.}
\lref\TT{A. Tyurin, ``{\it Fano versus Calabi-Yau},'' math.AG/0302101.}
\lref\Pol{J. Polchinski, `` {\it Dirichlet Branes and Ramond-Ramond
Charges},'' Phys.Rev.Lett 75 (1995) 4724.}
\lref\KMMM{Matrix models....}
\lref\Kap{M. Kapranov, ``{\it Rozansky-Witten invariants via Atiyah
    classes},'' alg-geom/9704009.}
\lref\SDstr{P.S. Howe, E. Sezgin and P.C. West, ``{\it
 The Six-Dimensional Self-Dual Tensors},'' Phys.Lett. {\bf B400} (1997)
255.}
\Title{ \vbox{\baselineskip12pt \hbox{hep-th/0409238} \hbox{HMI-04-01
}
\hbox{TCD-MATH-04-14}
\hbox{ITEP-TH-39/04} }} {\vbox{
\centerline{Towards Integrability of Topological Strings I:}
\vskip 1cm
\centerline{Three-forms on Calabi-Yau manifolds}
\bigskip
 }}
\medskip
\centerline{\bf Anton A. Gerasimov $^{1,2,3}$ and Samson L.
Shatashvili $^{2,3,4}$}
\vskip 0.5cm \centerline{\it $^{1}$ Institute for Theoretical and
Experimental Physics, Moscow, 117259, Russia} \centerline{\it
$^{2}$ Department of Pure and Applied Mathematics, Trinity
College, Dublin 2, Ireland } \centerline{\it $^3$ Hamilton
Mathematics Institute, TCD, Dublin 2, Ireland}
\centerline{\it $^{4}$ IHES, 35 route de Chartres,
Bures-sur-Yvette, FRANCE}
\vskip 1cm
The precise relation between Kodaira-Spencer path
integral and a particular wave function in seven dimensional
quadratic field theory is established. The special properties of
three-forms in 6d, as well as Hitchin's action
functional, play an important role.
The latter defines  a quantum field theory similar to
Polyakov's formulation of 2d gravity; the
curious analogy with world-sheet action of
bosonic string is also pointed out.
\medskip
\noindent \Date{}
\newsec{Introduction}
It has been suspected for long time that the  partition function of
type B topological strings on Calabi-Yau  manifold is related
to a wave function for some quadratic field theory in higher
dimensions. The most straightforward  realization of this idea
suggests the representation of the partition function
as a holomorphic wave function in the seven dimensional
theory with the phase space being the third
cohomology $H^3(M,\R)$ of the Calabi-Yau (CY) manifold $M$. In order to
be compared with the generating function of correlators in
topological string theory (partition function) this wave function should
be defined in
the linear polarization of the symplectic manifold $H^3(M,\R)$
associated with some reference complex structure on $M$. The
dependence on the reference holomorphic structure is governed by
the holomorphic anomaly equation \V,\BCOV,\W. However, even after the 7d
theory is
identified, the wave function is
not
unique and the problem of the construction of the
appropriate
wave function remains.

The  explicit  proposal
for the B-model target space field theory action was given in \BCOV,
where the generating function of correlators in the
topological theory  was represented as the path integral for
certain field theory on target space Calabi-Yau  manifold. The critical
points of this action correspond to the
solutions of the Kodaira-Spencer (KS) equation describing
the deformations of the complex structures on $M$.   However
the role played by this path integral in the above interpretation in terms
of the
holomorphic wave function was un-clear.

The obvious goal of all these studies is to find the proper, background
independent, target space
formulation for topological strings (target space string field theory)
which hopefully will be effective enough
to lead to the exact solution of the theory - integrability.

In this paper we establish the precise relation between the KS functional
integral
\BCOV\ and the wave function of the quadratic field
theory in seven dimensions (in the companion paper \GStwo\ we will
discuss the steps towards the integrability
based on lessons learned here). We study two natural
polarizations, linear and non-linear, in the phase space of 7d theory
and construct  integral
transformation connecting the wave functions in these two
polarizations. This result is exact, at
least in the quasi-classical approximation.
It appears that the path integral representation of
Kodaira-Spencer quantum field theory of \BCOV\ coincides with this
integral
transformation for a choice of a very simple wave function in the
non-linear
polarization.  This
non-linear polarization for the symplectic space $H^{3}(M,\R)$ is
constructed using the special properties of the three-forms in
six-dimensional space (see  \Hit,\Hitone,\Hittwo)
and  does not depend on any reference
complex structure on $M$. One can think about such construction as
a step towards
background independent formulation
of the theory \foot{Let us stress that the ambiguities of the quantization
in the non-linear polarization  might lead to the corrections  to KS
action
beyond the quasi-classical approximation, however the assumption of the
locality of the action in the natural coordinates may be used in order to
fix KS action unambiguously.}.

 It is
interesting to note that the same quadratic 7d theory
 turns out to be useful for the description of the
theory of self-dual three-forms in 6d. In this case
another linear polarization (based on Hodge $*$-operation)
is more adequate; for example the considerations in \SDstr,
\ref\WittenFive{E. Witten, ``{\it Five-Brane Effective Action In M-Theory},''
J.Geom.Phys. 22 (1997) 103-133, hep-th/9610234.} are close
to those presented below.

The relation between the wave function in 7d and 6d (``chiral") path
integral brings the
natural question - what is the interpretation of the
corresponding ``non-chiral counterpart" entering the scalar product of
the wave functions with the canonical measure? We briefly discuss this
question at the end of the paper and propose the interpretation via
the (modified) Hitchin functional - a possible six dimensional counterpart
of the Polyakov formulation for the 2d gravity where chiral part
(conformal
block) has well-known interpretation
in terms of wave-function in 3d Chern-Simons theory.

The plan of the paper is as follows. In Section 1 we start with the brief
description of the Kodaira-Spencer field theory formulated in
\BCOV; we rewrite the action in terms of the variables that will be useful
later. In Section 2 we quantize the 7d quadratic theory and
demonstrate the role played by KS path integral (written in above
variables)
as a wave function in the linear polarization, canonically related to a
simple
wave function in non-linear polarization.
In the Appendix 1 we present the
main, useful,
facts of the geometry of three-forms following Hitchin. In Appendix 2
some curious analogy between Nambu-Goto/Polyakov action in two
dimensions
and Hitchin functional introduced in \Hit\ is demonstrated. In Appendix 3
we give a simple derivation of the Holomorphic anomaly equation \V,\BCOV,
governing the dependence of the wave function in the linear
polarization on the base point of the moduli space. Our derivation is
based on the general properties of the special geometry of the moduli
space of
complex structures.

Some results
presented in this paper were known to the authors for a period
of time, originating to mid 90's, and were  presented in the
talks on various occasions
\SS, \AG. We are grateful to M. Kontsevich, G.
Moore, N.
Nekrasov, C. Vafa, E. Verlinde and E. Witten for discussions of
related issues over the period of time since 1993. The research of
A. A. Gerasimov is partly supported by the
 grant RFBR 03-02-17554 and SFI Basic Research Grant;
 that of S. L. Shatashvili by SFI Basic Research Grant.
\newsec{Geometry}
\subsec{Kodaira-Spencer Theory} In this section we recall some facts about
Kodaira-Spencer theory following \BCOV\ and rewrite the action in terms of
the variables appropriate for further considerations.

Let
$M$ be a compact Calabi-Yau (CY) manifold. Gauged CY manifold is a pair
$(M,\Omega_0)$ where $M$ is a   CY manifold supplied with a
holomorphic $(3,0)$-form $\Omega_0$. Holomorphic $(3,0)$-form on
$M$  is uniquely defined up to the multiplication by a  non-zero
complex number and thus the moduli space $\widehat{{\cal M}}_M$
of the gauged CY manifolds is a $\C^*$-bundle over the moduli space
${\cal M}_M$ of complex  structures on $M$. Fixing particular
holomorphic $(3,0)$-form $\Omega_0$ defines, locally, a section of
the bundle. In the vicinity of the point of the maximal
degeneration of the complex structures  the existence of the
weight filtration on the homology leads to the natural choice of
the three-cycle $\gamma_0 \in H_3(M,\Z)$ and thus to the natural
normalization condition  $\int_{\gamma_0}\Omega_0=1$ \DM.  Another way
to normalize the holomorphic three-form is to use the condition
$\int_M \Omega_0\wedge (\overline{\Omega_0})_m=1$ where
$(\Omega_0)_m$ is a fixed holomorphic $(3,0)$-form for the
reference complex structure $m\in {\cal M}_M$.

 The choice of the complex structure provides
the  decomposition of the exterior derivative:
$D=D^{1,0}+D^{0,1}=\pr +\apr$.  Below we will use the
notation: $\Omega^{-p,q}(M)\equiv \Omega^q(M,\wedge^p
{T})$, where ${T}$ is a holomorphic tangent bundle.

Let  $(z^i,\az^{\ai})$ be  local coordinates on $M$ and let  $\AAA
\in \Omega^{-1,1}(M)$ be a $(-1,1)$-differential
 written locally as: $\AAA=\sum \AAA_{\ai}^jd\az^i\frac{\pr}{\pr
z^j}$. Then the  deformation of the complex structure may be
described in terms of the deformation of the operator
$D^{0,1}=\apr$ :
\eqn\two{
 \apr \rightarrow \apr_{\AAA}=\apr
 +\AAA=\sum d\z^{i}(\frac{\pr}{\pr
   \az^i}+\AAA_{\ai}^j\frac{\pr}{\pr z^j}),}
 subjected to the  integrability condition $\apr_{\AAA}^2=0$
(Kodaira-Spencer equation).
 This is equivalent to the following equation:
 \eqn\KS{
\apr \AAA+\frac{1}{2}[\AAA,\AAA]=0,} or explicitly
\eqn\ksone{(\apr_{\ai} \AAA^k_{\aj}+\AAA^l_{\ai}\pr_l
              \AAA^k_{\aj})d\az^{i}\wedge d\az^{j}\frac{\pr}{\pr z^k}=0.}
The moduli space of complex structures is given by the space of
the solutions of the equation \KS\  modulo the gauge
transformations \eqn\gaugetr{
 \delta \AAA=\apr_{\AAA}\epsilon =\apr \epsilon
 +[\AAA,\epsilon],}
 where $\epsilon \in \Omega^{-1,0}(M)$.
Equivalently the moduli space of the complex structures may be
parameterized by the solutions of the pair of the equations:
\eqn\KSful{\eqalign{& \apr \AAA+\frac{1}{2}[ \AAA,\AAA ]=0, \cr &
\pr A=0,}} modulo the subgroup of the gauge transformations
\gaugetr\  which leave the holomorphic three-form $\Omega_0$
invariant.
Note that \KSful\ is  equivalent to the
deformation of the pair of the operators $(\apr,\pr)$ :
\eqn\three{\eqalign{& \apr \rightarrow \apr +\AAA, \cr & \pr
\rightarrow \pr,}}
with the relations: \eqn\commut{\eqalign{
 \apr_{\AAA}^2=\pr^2=\apr_{\AAA} \pr +\pr \apr_{\AAA}=0}.}

Given a holomorphic  $(3,0)$-form $\Omega_0$ we could identify
$\Omega^{-p,q}(M)$  with $(3-q,p)$-forms $\Omega^{3-p,q}(M)$.  The
interior product $v\vdash \omega$ of the vector field $v\in
Vect_M$ and the differential $n$-form $\omega$ is defined as:
$(v\vdash \omega)=\sum_{a_1\ldots a_n}  (-1)^k
v^{a_k}_k\omega_{a_1\ldots a_k \ldots
  a_n}dx^{a_1}\wedge \ldots \widehat{dx}^{a_k} \ldots \wedge
dx^{a_n}$. Extending this operation to the polyvector fields  we
obtain  the map:
 \eqn\four{\eqalign{&
 \Omega^{-p,q}(M) \rightarrow
 \Omega^{3-p,q}(M),\cr
& A \rightarrow  A^{\vee}= A \vdash \Omega_0. }}
Under this
identification the equation \KSful\ reads:
\eqn\constroneone{
\apr  \AAA^{\vee}+\frac{1}{2}\pr( \AAA \wedge  \AAA)^{\vee}=0,}
\eqn\constronetwo{
\pr \AAA^{\vee}=0,}
and the gauge transformation \gaugetr\ is:
$$
\delta \AAA^{\vee}=\apr\, \epsilon^{\vee}+\pr(\AAA \wedge
\epsilon)^{\vee},
$$
where $\epsilon^{\vee}$ is constrained by the condition $\pr
\epsilon^{\vee}=0$.

Let us define  the following
 operations  on the  forms:
$$
 A^{\vee}\circ  B^{\vee}= (A\wedge B)\vdash
 \Omega_0,$$
 $$
 <A^{\vee},B^{\vee},C^{\vee}>= A^{\vee}\wedge(B^{\vee}\circ C^{\vee}),
$$
 where $A^{\vee}=A\vdash \Omega_0$, $B^{\vee}=B\vdash \Omega_0$,
 $C^{\vee}=C\vdash \Omega_0$.
Then the system of the equations \KSful\ becomes equivalent to:
\eqn\eight{\eqalign{& \apr \AAA^{\vee}+\frac{1}{2}\pr(
\AAA^{\vee}\circ \AAA^{\vee})=0 \cr
 & \pr \AAA^{\vee}=0.}}
From the second equation in the case of the compact K\"ahler
manifold one has: \eqn\param{
 \AAA^{\vee}=x+\pr b,}
where $x$ is a $\pr$-harmonic $(2,1)$-form and $b \in
\Omega^{1,1}(M)$. Now the equation \eight\ becomes the equation for $b$:
 \eqn\eqphi{ \apr \pr b+\frac{1}{2 }\pr((x+\pr b)\circ (x+\pr
b))=0.}
We note that this equation has a meaning of anti-holomorphicity for
following (1,2)-current:
\eqn\new{{\bar J}^{(1,2)}=\apr b+\frac{1}{2 }(x+\pr b)\circ (x+\pr
b); \quad \quad \pr {\bar J}^{(1,2)}=0.}

The action functional leading to the equations of motion in the
 form \eqphi\ is \eqn\KSaction{ S_{KS}(b|x)=\int_M
(\frac{1}{2}\pr b \wedge \apr b+\frac{1}{6}<(x+\pr
 b),(x+\pr b),(x+\pr b)>).}
This action coincides with the action \KS\  introduced in
\BCOV.
 \eqn\BCOVact{
S_{KS}'(\AAA^{\vee})=\int_M (\frac{1}{2} \AAA^{\vee} \apr
\,\pr^{-1} \AAA^{\vee}+\frac{1}{6}
<\AAA^{\vee},\AAA^{\vee},\AAA^{\vee}>).}
In order to get precise relation one should  solve the constraint
\constronetwo\
and use parameterization \param\ to resolve
the non-localities in the first term  in \BCOVact.

\subsec{KS
theory in terms of  7d quadratic field theory}

Now we are ready to show that the functional integral with the action
\KSaction\
\eqn\KSTWENTYFIVE{ Z(x)=\int Db \,\,e^{-S_{KS}(b|x)},}
may be considered as an integral representation of some particular  wave
function in  the seven-dimensional  field theory on $M\times \R$
 (the pruduct of $M$ and the real line $\R$)
with the action functional:
\eqn\nine{ S(C)=\int_{M\times \R} C d C.}
Here  $C$  is a real three-form  and the corresponding quantum field
theory may be considered as a higher dimensional generalization of
the abelian Chern-Simons theory in three dimensions. Written in
$6+1$ notations in local  coordinates $(x^i,t)$ on  $M\times
\R$ one has:
\eqn\ten{ S=\int_{M\times \R} dt\, d^6x\, (\Omega
\frac{\pr}{\pr t} \Omega+\omega_t d\Omega),} where
 $C=\Omega+\omega_tdt$ with  $\Omega$ -  a
three-form component of $C$  along $M$, $\omega_t \,dt$ - a
two-form along $M$ and one-form along $\R$.

Consider the infinite-dimensional space of real three-forms on $M$  supplied with the
 symplectic structure:
 \eqn\eleven{
\omega^{symp}(\delta_1 \Omega, \delta_2 \Omega) =\int_M \delta_1
\Omega \wedge \delta_2 \Omega.}
The phase space for \nine\ is obtained  by applying the Hamiltonian reduction
via imposing the first class constraint:
 \eqn\constrtwelve{ d\Omega=0.}
In order to construct the wave function explicitly one needs to
chose a polarization. After the polarization is chosen - the constraints
are  imposed  by   demanding  that the wave function is a solution
of the system of differential equations, following from \constrtwelve.
In the simple cases the formal
solution may be
given in the integral form by applying the appropriate averaging
procedure.

There are several natural polarizations on the
space of three-forms in six-dimensions defined in terms of the
families of the Lagrangian sub-manifolds. Below  two polarizations, one of
which is
linear and another one is non-linear will be important.
 It appears that KS partition function  \KSTWENTYFIVE\ is naturally
connected with wave function  in the linear polarization. On the
other hand the most natural expression for the wave function is in
the non-linear polarization. Thus we start with the simple
expression for the wave function in the non-linear polarization
and then transform it to the linear polarization using the appropriate
unitary transformation. Finally we construct the wave function in
the constrained theory by imposing the constraints (using the
averaging procedure).

Let us start with the description of the
polarizations we will use. The simplest  one is a linear
polarization. Given a complex structure on $M$  one could
decompose the space of complex three-forms $\Omega_{\C}$ over
  the forms of the definite Hodge type:
 \eqn\thirteen{
 \Omega_{\C} =\Omega^{3,0}\oplus \Omega^{2,1} \oplus \Omega^{1,2}
 \oplus \Omega^{0,3}.}
The real forms are singled out by the reality condition:
$\Omega^{0,3}=\overline{\Omega^{0,3}}$,
 $\Omega^{1,2}=\overline{\Omega^{2,1}}$.

The subspace
 $\Omega^{3,0}\oplus \Omega^{2,1}$ defines the complex
Lagrangian (linear) sub-manifold in the space of complex three-forms
and thus the complex polarization of the space of real
three-forms.
 It will be useful to parameterize this  subspace  as
 \eqn\poonetwothree{
 \Omega_c=\rho(\Omega_0+\overline{A}\vdash \Omega_0),}
 where $\Omega_0$ is a fixed normalized  holomorphic $(3,0)$-form
in a reference  complex structure, $\rho$ is a function on $M$  and
  $\overline{A}$ is a $(-1,1)$-differential.

 In the specific case at hand of the three-forms
in 6d there is another polarization. It
 may be constructed  using  the decomposition of the general real
 three-forms $\Omega$ into the sum of two decomposable forms:
 $\Omega=\Omega_++\Omega_-$.  We say that the
 real three-form is non-degenerate if the top-dimensional  form
$\Omega_+\wedge
 \Omega_-$ has no zeros. The space of the non-degenerate   three-forms
consists of two non-intersecting parts $U_+\cup U_-$ depending  on
 whether the decomposable forms $\Omega_{\pm}$ are real - $\Omega \in
U_+$,
or complex
 conjugate to each other - $\Omega \in U_-$ (see the precise description
in the
 Appendix 1).

 Let us consider  the forms  in $U_-$ having the following decomposition
\eqn\split{
 \Omega=\Omega_++\Omega_-=E^1\wedge E^2\wedge E^3+\overline{E}^1\wedge
 \overline{E}^2\wedge \overline{E}^3,}
with  $E^i$  being complex one-forms.  $(E^i,\overline{E}^i)$
generically produces the frame in the complexified cotangent
bundle $T_{\C}^*M$  and
 $\Omega_-\wedge \Omega_+\neq 0$. The subspace
of the decomposable forms  defines  the Lagrangian family:
\eqn\fourteen{
 \omega^{symp}(\delta_1 \Omega_+ ,\delta_2 \Omega_+)=\int
 \delta_1 (E^1\wedge E^2\wedge E^3)  \wedge
\delta_2(E^1\wedge E^2\wedge E^3)=0.} Locally the decomposable
three-form $\Omega_+$
 may be parameterized as:
 \eqn\decomp{\eqalign{&
 \Omega_+=\frac{1}{6}\epsilon_{ijk} \varrho
(dz^i+\mu^i_{\ai}d\az^{\ai})(dz^j+\mu^j_{\aj}d\az^{\aj})
(dz^k+\mu^i_{\ak}d\az^{\ak})=\cr
   &= \varrho (\Omega_0+\mu \vdash \Omega_0+\frac{1}{2}\mu^2\vdash
   \Omega_0+\frac{1}{6}\mu^3\vdash \Omega_0)=\varrho e^{\mu
 \vdash}\Omega_0},}
where  $\mu \in \Omega^{-1,1}(M)$ and $\varrho$ is a function
 on $M$ and we use the notations $\mu^n\vdash := (\mu \vdash)^n$.

\subsec{Linear polarization}

Let us proceed  with the description
of the wave functions in the linear
  polarization \foot{We abuse the notations by using
  $\Omega^{p,q}$ for the space of the forms of the given Hodge type
  and for particular $(p,q)$-differential form.}.
In this polarization the constraint equations $d\Omega=0$:
$$\apr \Omega^{3,0}+\pr \Omega^{2,1}=0,$$
 $$\apr \Omega^{2,1}+\pr \Omega^{1,2}=0,$$
 $$\apr \Omega^{1,2}+\pr \Omega^{0,3}=0,$$
quantum mechanically take the form of the following equations on the wave
function:
$\Psi(\Omega^{3,0},\Omega^{2,1})$
 \eqn\constrfiftysix{\eqalign{&
( \apr \Omega^{3,0}+\pr \Omega^{2,1})\Psi=0, \cr
 &(\apr \Omega^{2,1}+\pr \frac{\delta}{\delta \Omega^{2,1}})\Psi=0, \cr
 &( \apr \frac{\delta}{\delta \Omega^{2,1}}+\pr \frac{\delta}{\delta
\Omega^{3,0}})\Psi=0.}}
The  formal solution of the constraints \constrfiftysix\ may
be written in terms of the path integral. Given an arbitrary
function $\Psi_0(\Omega^{3,0},\Omega^{2,1})$, one can construct the
formal solution    representing it in
the form:
\eqn\intzero{
\Psi(\Omega^{3,0},\Omega^{2,1})=(\Pi \Psi_0)(\Omega^{3,0},\Omega^{2,1}),}
where the projection operator $\Pi$ is given by:
 \eqn\intone{\Pi=\int D\Lambda\, D\sigma \, Db \,\,
e^{\int_M \Lambda (\apr \Omega^{3,0}+\pr \Omega^{2,1})} e^{\int_M
\sigma (\apr \frac{\delta}{\delta \Omega^{2,1}}+ \pr
    \frac{\delta}{\delta \Omega^{3,0}})}  e^{\int_M b(\apr
    \Omega^{2,1}+\pr \frac{\delta}{\delta \Omega^{2,1}})}.}
  The only restriction on
  $\Psi_0$ is the convergence of the integrals in \intone.
We can use the identity:  $\exp(A+B)=\exp(A-\frac{1}{2}[A,B])\exp(B)$
for $A,B$ such that: $[A,[A,B]]=[B[A,B]]=0$ and get:
$$
\Psi(\Omega^{3,0},\Omega^{2,1})=\int D\Lambda \,D\rho\, Db\,\,
e^{\int_M \Lambda (\apr \Omega^{3,0}+\pr \Omega^{2,1})+
 \frac{1}{2}\pr b \apr b+\apr b\wedge \Omega^{2,1}}
\Psi_0(\Omega^{3,0}-\pr \sigma,\Omega^{2,1}-\pr b-\apr \sigma).$$
In order to simplify this representation let us use the
 following parametrization of $(\Omega^{3,0},\Omega^{2,1})$ in
terms of the non-zero constant $\rho_0$, $\,(2,0)$-form $\chi$ and
$(2,1)$-form $\widetilde{\Omega}^{2,1}$
 \eqn\sixteen{\eqalign{ &
\Omega^{3,0}=\rho_0\Omega_0+\pr \chi, \cr
&\Omega^{2,1}=\widetilde{\Omega}^{2,1}-\apr \chi.}}
In these variables:
\eqn\seventeen{
\Psi(\Omega^{3,0},\Omega^{2,1})=\Psi(\rho_0,\chi,\widetilde{\Omega}^{2,1})=\int
D\Lambda\, D\sigma \, Db\,\, \times }
$$\times e^{\int_M \Lambda
(\pr\widetilde{\Omega}^{2,1})}
 e^{-\int_M(\frac{1}{2}\pr b \apr b+\apr b\wedge
\widetilde{\Omega}^{2,1})} \Psi_0(\rho_0 \Omega_0-\pr \sigma
,\widetilde{\Omega}^{2,1}-\pr b-\apr
  \sigma ).$$
Any  form $\widetilde{\Omega}^{2,1}$  has $\pr$-Hodge
decomposition:
 \eqn\eighteen{
 \widetilde{\Omega}^{2,1}=x+\pr\lambda +\pr^{\dagger}\widetilde{\lambda},}
with  $x$ being  a harmonic form ($\pr x=\pr^{\dagger} x=0$). We can use
this representation and replace the integral over $\Lambda$ by the
functional
  delta-function, so we finally have:
  \eqn\ninteen{\eqalign{& \quad \quad \quad \quad \quad
\Psi(\Omega^{3,0},\Omega^{2,1})=\Psi(\rho_0,\chi,x,\lambda,\widetilde{\lambda})=\delta(\pr
\pr^{\dagger}\widetilde{\lambda}) \times \cr &
\times e^{-\int_M(\frac{1}{2}\pr
\lambda \apr \lambda)} \int D\sigma  \,Db\,\,
e^{-\int_M(\frac{1}{2}\pr b \apr b+\apr b\wedge \pr \lambda)}
\Psi_0(\rho_0\Omega_0-\pr \sigma ,x-\pr b-\apr \sigma ).}}

The
symplectic structure is constant in terms of the variables
$\Omega^{p,q}$ and the scalar product of the wave functions in
given by:
\eqn\twenty{\eqalign{ <\Psi_1|\Psi_2>=\int
D\Omega^{3,0}\, & D\Omega^{2,1}\, D\Omega^{1,2}\, D\Omega^{0,3}\,\,
e^{\int_M (\Omega^{3,0}\wedge \Omega^{0,3}+ \Omega^{2,1}\wedge
\Omega^{1,2})}\times \cr  & \times
\overline{\Psi}_1(\Omega^{0,3},\Omega^{1,2})\Psi_2(\Omega^{3,0},\Omega^{2,1}).}}
where integration is over the real subspace
$
\overline{\Omega^{0,3}}=\Omega^{3,0}$, $\overline{
\Omega^{1,2}}=\Omega^{2,1}$.

Such representation of the wave function in the linear
polarization may be interpreted in the same fashion as the  relation
between 3d Chern-Simons theory and 2d CFT - the wave function is
the deformed partition function in the 6d quadratic field theory
of two-forms. The examples of the interesting wave functions are given
by the appropriate chiral deformations of the functional integrals
in quadratic theories (i.e $(B,C)$-systems) in 6d (holomorphic equation
\new\ can serve as the starting point for such consideration). Below we will be
interested in the particular wave function described in terms of the
KS path integral.

 \subsec{Nonlinear polarization}

 Now we consider the
quantization in  the non-linear polarization \split, \decomp:
\eqn\polone{\eqalign{&
 \Omega_{+}=\frac{1}{6}\epsilon^{ijk}
 \varrho (dz^i+\mu^i_{\ai}d\az^{\ai})(dz^j+\mu^j_{\aj}d\az^{\aj})
(dz^k+\mu^i_{\ak}d\az^{\ak})=\cr &= \varrho(\Omega_0+\mu\vdash
\Omega_0+\frac{1}{2}\mu^2\vdash
   \Omega_0+\frac{1}{6}\mu^3\vdash \Omega_0),}}
and \eqn\twentyone{
 \Omega_{-}=\overline{\Omega}_+=
\overline{\varrho}(\overline{\Omega}_0+\overline{\mu}\vdash
   \overline{\Omega}_0+\frac{1}{2}\overline{\mu}^2\vdash
  \overline{\Omega}_0+\frac{1}{6}\overline{\mu}^3\vdash
\overline{\Omega}_0).}
We realize the quantum state as a function of $(\varrho,\mu)$ variables.

In order to describe
the scalar product in this polarization we need to find the generating
function
$S_0(\varrho,\mu,\overline{\varrho},\overline{\mu})$ of the
canonical transformation  from a set of variables $(\varrho,\mu)$ to
a set of complex conjugate variables
$(\overline{\varrho},\overline{\mu})$. This can be done by use of
the standard expression for the  generating function
$S(Q,q)$ for canonical transformation from $(p_i,q^i)$ to $(P_i,Q^i)$:
$\sum_i P_i \delta Q^i-\sum_i
p_i \delta q^i=\delta S(Q,q)$:
\eqn\newone{ \int_M
\Omega_-(\overline{\varrho},\overline{\mu})\delta
\Omega_+(\varrho,\mu)- \int_M \Omega_+(\varrho,\mu)\delta
\Omega_-(\overline{\varrho},\overline{\mu})=\delta \int_M
(\Omega_-(\overline{\varrho}, \overline{\mu})\wedge
\Omega_+(\varrho,\mu)),} and therefore  the scalar product is given
by \eqn\twentyfour{ <\Psi_1|\Psi_2>= \int
{\cal D}(\mu,\overline{\mu},\varrho,\overline{\varrho})e^{\int_M
  \Omega_-(\overline{\varrho},
\overline{\mu})\wedge \Omega_+(\varrho,\mu)}
\overline{\Psi}_1(\overline{\varrho},\overline{\mu})\Psi_2(\varrho,\mu).}
The exponential factor in the integration measure may be
 written in terms of the Hitchin functional $\Omega_+(\Omega)\wedge
\Omega_-(\Omega)=-\sqrt{\l(\Omega)}$ (see Appendix 1).

Imposing the constraints in
this polarization  however is note quite trivial procedure. The reason for
this is
that the gauge  symmetry:
 \eqn\twentyone{
 \Omega \rightarrow \Omega +d\xi,}
generated by constraints does not respect the polarization.
The decomposition of the gauge transformed three-form $\Omega$ on
the decomposable parts: \eqn\twentytwo{
\Omega=\Omega_++\Omega_-\rightarrow \Omega+d\phi=(\Omega_++\delta
\Omega_+(\Omega_{\pm},\phi)) + (\Omega_-+\delta
\Omega_-(\Omega_{\pm},\phi)),} is  highly  non-linear in terms of
the initial $\Omega_{\pm}$ (see Appendix 1). Thus in this polarization
the gauge  transformation  mixes ``coordinates'' and ``momenta''
 in a complicated  way and  the constraints are  given by
rather complex differential operators acting on the wave
function. Therefore  we will use the following strategy: start with the
simple,
unconstrained wave-function in the non-linear polarization $\rightarrow$
transform this wave function into the corresponding wave function
 in the linear polarization $\rightarrow$ impose the constraints. This
gives us
 the constraint wave function in the linear polarization whose
 particular form  reflects the simplicity of the initial wave function in
the
  non-linear polarization.

Let us start with the explicit transformation for the wave functions
between linear and non-linear polarizations.
From
\poonetwothree\ and \polone\ we establish the relations:
\eqn\transone{\rho\Omega_0=\varrho\Omega_0+\frac{1}{6}
\overline{\varrho}\,\overline{\mu}^3\vdash\overline{\Omega}_0,}
\eqn\transtwo{\rho \AAA \vdash \Omega_0= \varrho \mu\vdash
 \Omega_0+\frac{1}{2}\overline{\varrho}\,
\overline{\mu}^2\vdash \overline{\Omega}_0,}
\eqn\trnasthree{\overline{\rho}A \vdash \overline{\Omega}_0=
\overline{\varrho}
\,\overline{\mu}\vdash
 \overline{\Omega}_0+\varrho\frac{1}{2}\mu^2\vdash \Omega_0,}
\eqn\transfour{\overline{\rho}\,\overline{\Omega}_0=
\overline{\varrho}\,\overline{\Omega} _0+\frac{1}{6}\varrho\mu^3
\vdash\Omega_0.} As coordinates (versus momenta) in  the linear
polarization  \poonetwothree\ we  use: $(\overline{\rho},A)$,
and in the non-linear polarization: $(\mu,\varrho)$.
The reason for this particular choice is the following. It is clear
that \transone, \transtwo, \trnasthree\ and \transfour\ may be
considered as  perturbations  of the trivial transformation by
the non-linear terms. But the kernel of the trivial transformation is
given by delta-function that could not be easily described
in the classical approximation as an  exponent of some
generating function. With the proposed choice of the polarization
there is no such problem because the  unperturbed transformation has
smooth kernel. Thus, we can rewrite the relations in \transone\ -
\transfour\
in the following form:
$$
 \overline{\varrho}=\overline{\rho}-\varrho <\mu^3>,$$
$$\overline{\mu}= \frac{A-\varrho
\overline{\rho}^{-1}\overline{\mu}^{\vee}}
{1-\varrho\overline{\rho}^{-1}<\overline{\mu}^3>},$$
$$\rho =\varrho +\overline{\rho }
\frac{<(A-\varrho \overline{\rho}^{-1}\overline{\mu}^{\vee})^3>}
{(1-\varrho \overline{\rho}^{-1}<\overline{\mu}^3>)^2},$$
$$\rho \AAA=\varrho \mu+
\overline{\rho} \frac{(A-\varrho
\overline{\rho}^{-1}\overline{\mu}^{\vee})^2} {(1-\varrho
\overline{\rho}^{-1}<\overline{\mu}^3>)},
$$
where the functions $<\mu^3>$, $<\overline{\mu}^3>$ and $(1,-1)$-form
$\overline{\mu}^{\vee}$ are defined by:
$$
<\overline{\mu}^3>\overline{\Omega}_0\wedge \Omega_0
=\frac{1}{6}\overline{\Omega}_0\wedge (\overline{\mu}^3\vdash
\overline{\Omega}_0)
$$
$$
<\mu^3> \Omega_0\wedge \overline{\Omega}_0 =\frac{1}{6}\Omega_0 \wedge
(\mu^3 \vdash \Omega_0), $$
$$\overline{\mu}^{\vee}\vdash \overline{\Omega}_0
=\frac{1}{2}\mu^2\vdash\Omega_0.$$

Now for the
generating function $S(\mu,
 \varrho,\overline{\rho},A)$ we have:
 \eqn\twentyeight{\eqalign{&  \quad \quad \quad \quad \quad \quad \quad
\quad \quad \quad
 \delta S(\mu,
\varrho,\overline{\rho},A)= \cr &
= \Theta_{\varrho}(\mu,\overline{\mu},\varrho,\overline{\varrho})\delta
 \varrho +
 \Theta_{\mu}(\mu,\overline{\mu},\varrho, \overline{\varrho})\delta \mu-
\Theta'_{\overline{\rho}}(A,\AAA,\rho,\overline{\rho})\delta
\overline{\rho} -\Theta'_{A}
 (A,\AAA,\rho,\overline{\rho})\delta A,}}
 where:
$$
\delta \Theta=\delta \Theta ' =\omega^{symp},$$
$$
\Theta'=\int_M \,
 \rho (\Omega_0+\AAA\vdash\Omega_0)\wedge
\delta(\overline{\rho}(\overline{\Omega_0}+A\vdash
\overline{\Omega}_0)), $$
$$\Theta= \int_M \Omega_-(\overline{\varrho},\overline{\mu})\wedge \delta
\Omega_ +(\varrho,\mu).$$
Solution is:
\eqn\thirtytwo{ \eqalign{& \quad \quad \quad \quad \quad \quad \quad \quad
\quad \quad \quad \quad
S(A,\overline{\rho},\mu,\varrho)= \cr & = \int_M
((\overline{\rho}\varrho+\varrho^2<\mu^3>+\frac{ <(A
\overline{\rho}-\varrho
\overline{\mu}^{\vee})^3>}{(\overline{\rho}-\varrho <\mu^3>)})
\Omega_0 \wedge
\overline{\Omega}_0+\overline{\rho}\varrho (\mu \vdash \Omega_0)
\wedge (A\vdash \overline{\Omega}_0)).}}
We conclude that
(at least in the quasi-classical approximation) the connection
between the wave functions in the linear and the non-linear
polarizations is given by:
\eqn\transform{\eqalign{
\Psi(\varrho,\mu)=&\int
D\overline{\rho}\,DA\,\,e^{S(A,\overline{\rho},\mu,\varrho)}\Psi(\overline{\rho},A),
\cr & \Psi(\overline{\rho},A)=\int
D\overline{\rho}\,DA\,\,e^{-S(A,\overline{\rho},\mu,\varrho)}\Psi(\varrho,\mu)
.}} This concludes the construction of the general wave
function in the unconstrained system in the non-linear
polarization. We emphasize that this representation is unambiguous only in
the semi-classical
approximation  and thus the meaningful part of the integration is
given by the evaluations at the critical points and determinant of
quadratic fluctuations around them.

In order to get the constrained wave function we
  transform this function into the wave function in
the linear polarization and impose the constraint.
This is equivalent to the explicit calculation of  the following matrix
element:
$$
\Psi(\Omega^{3,0},\Omega^{2,1})=<\Omega^{2,1},\Omega^{3,0}|\Pi|\psi>,$$
where $<\Omega^{2,1},\Omega^{3,0}|$ is the coordinate eigenvector
in the linear polarization,  $\Pi$ is the projector imposing the
constraints and $|\psi>$ is some state that we will now explicitly define
in the non-linear polarization.

We claim that the following choice leads to desired result:
\eqn\testfun{
 \psi(\overline{\varrho},\overline{\mu})=\delta(\overline{\mu})\exp \int_M
\overline{\varrho}.}
The reasoning under such selection is following: it is
 easy to see that in the discussed approximation the wave function in
 the $(\varrho,\mu)$-polarization,  corresponding to \testfun,
 is given  by:
\eqn\testfunone{
 \psi(\varrho,\mu)=\delta(\varrho-1).}
This corresponds to the fixing of the holomorphic volume form
 (stated differently - to the choice of the closed string coupling
 constant).
Thus we have:
$$
 \Psi(\Omega^{3,0},\Omega^{2,1})=\int D\overline{\mu}
D\overline{\varrho}\,\,e^{-S(\Omega^{3,0},\Omega^{2,1},\overline{\mu},\overline{\varrho})}
\delta(\overline{\mu})\exp \int_M \overline{\varrho}=$$
$$ =\delta(\Omega^{3,0}- \Omega_0)\exp(-\int_M
\frac{1}{6}<\Omega^{2,1},\Omega^{2,1},\Omega^{2,1}>).
$$
Finally, the action of the projection operator gives:
\eqn\thirtyfour{\eqalign{&\Psi(\Omega^{3,0},\Omega^{2,1})=
\Psi(\rho_0,\chi,x,\lambda,\widetilde{\lambda})=\delta(\rho_0-1)
\delta(\pr\pr^{\dagger}\widetilde{\lambda})
 e^{-\int(\frac{1}{2}\pr \lambda \apr \lambda)} \,
 \times \cr & \times
 \int  Db \,\,
e^{-\int(\frac{1}{2}\pr b \apr b+\apr b\wedge \pr
  \lambda+ \frac{1}{6}<(x+\pr b),(x+\pr b),(x+\pr b)>)}.}}
 Specializing to the subspace $\lambda=0$ we obtain,  up to
the $x$-independent factor, the following integral representation for
the wave function:
\eqn\thirtyfive{\Psi(\Omega^{3,0},\Omega^{2,1}) \rightarrow
 \Psi(x) = const\, \int  Db \,\, e^{-\int(\frac{1}{2}\pr b \apr
  b+\frac{1}{6 }<(x+\pr b),(x+\pr b),(x+\pr b)>)}.}
This representation coincides with the KS quantum field theory path
integral  \KSaction.

\newsec{Conclusions and further directions}

The derivation of Kodaira-Spencer path integral presented above seems
rather  appealing  from the physical point of view. This approach  is
based
on
the use of the vacuum wave function in the first quantized formalism
(holomorphic three-form) as a fundamental variable; the connection
with the seven-dimensional formulation  bears obvious similarity with
2d-gravity/3d Chern-Simons relation \Wtwo, \HV. \foot{
Let us remark that  the formulations of the gravitational theories in
terms of the gauge theory (if any) usually has the  problem with the
proper
definition of the configuration space. The
condition of the positivity of the metric removes some region in the
configuration space of the gauge  fields. Moreover the factorization
over the mapping class group $Diff(M)^+/Diff_0$
 is usually added by hand \Wtwo.  Nevertheless such representations may
turn out to be useful in
the search of the exact solution.}

The appearance of Hitchin's action functional as the gluing factor in
\newone\
is not accidental (this action functional is written in Appendix 1 in an
invariant form (4.3),
where we also list some of its important properties;
see also Appendix 2). In above-mentioned analogy, 2d gravity/3d
Chern-Simons relation,
the Hitchin action plays the role of Liouville theory and KS path integral
- Virasoro conformal block. In this respect Hitchin action can be viewed as
an interesting candidate for an
alternative definition of target space quantum theory.
One of the nice properties of this action is that it is almost
background independent - one only needs to fix the class of
three-form in $H^3(M,\R)$.

In conclusion we would like to list some obvious further directions of
study.
The most obvious one is to incorporate the generalized
complex structures introduced by Hitchin in \Hthree.  The generalized
complex
structures are described in terms of the families of  the Lagrangian
subspaces in
the fibers of the direct sum of the tangent and cotangent bundles
$TM\oplus T^*M$. From the seven-dimensional perspective it turns out to be
equivalent to the replacement of the three-form $C$ by an arbitrary odd
form of mixed
degree with  the same quadratic action as used above. The
appearance of the $\Z_2$-grading instead of $\Z$-grading
strongly suggests the connection  with the
K-theory and the formulation in terms of the Dirac operator. Corresponding Hitchin
action should be viewed as a non-chiral completion of "chiral" theory corresponding
 to KS action.

Special properties of three-forms in six dimensions are not unique -
three-forms and four-forms in seven and eight dimensions have very
similar behavior. There is an evidence in the favor of the relation
between these higher dimensional theories (of three and four-forms)
and the topological strings. It has been conjectured previously \SS\
that topologically twisted $G_2$ model
(N=1 Superconfromal theory with 7d target \ref\VS{S. Shatashvili and C. Vafa,
{\it Superstrings and Manifolds of Exceptional Holonomy,} hep-th/9407025, Selecta Mathematica,
New Series, Vol. 1, No. 2 (1995) 347 - 381.})
shall describe both the deformation of the
Complex and K\"ahler (both A and B model) structures for the CY
manifold sitting inside $G_2$-manifold simultaneously, though precise realization was not known
(see the last section in \VS\ in regard to topological twist in $G_2$
superconformal model on world-sheet).

The other obvious direction is  the search for the
KS type theories for the more general
 (higher-dimensional, non-compact, ...)
 CY manifolds and even more abstract non-geometric backgrounds. In this
respect one
shall stress the crucial role of the decomposability
 condition used in this paper which is closely tied to
 the criticality of 6d. One the other side, for example the
 Pl\"{u}cker-Hirota equations in KP theory are
just the conditions of the decomposability of the semi-infinite forms and
one would wonder if
such systems come to play for non-geometric and other backgrounds?

Probably the most intriguing application is related to the
possible reformulation of the critical string theory,
especially in the view of the fact
that KS theory is a kind of
string field theory \BCOV.
The
lesson we learned is that the holomorphic volume form is the natural
variable for background independent formulation (the choice of
correct variables is
a key ingredient in finding the exact solution of such systems, or in
establishing gravity/gauge theory relations)
\foot{According to \DT\ this means that the dynamical variable
is a complex
    orientation of the six dimensional manifold. }. More
    abstractly - the holomorphic volume form in KS theory  may be
    considered as a wave
    function in the first quantized formalism. This interpretation
implies a straightforward generalizations (compare with the similar
approach to reformulating critical string theory in \BR). However it seems that
  holomorphic volume form
 could hardly be considered as a description of truly fundamental degree of
  freedom. In this respect we recall other important
  lesson learned during the study of the non-critical strings in late 80's
- for those,
 rather special  cases,   the most natural formulation was obtained in
terms
of the 2d
  fermions. The important step was
the construction proposed by Kontsevich  \Kon. It  can be reformulated in terms of the
simple quadratic fermionic functional integral
\AG\ thus leading to the explicit description of the large
class of the topological B models  (to be compared with  the old results, see for example
\ref\GMoore{P. Ginsparg and G. Moore, {\it Lectures on 2D gravity and 2D
string theory,} (TASI 1992), hep-th/9304011.},
recent progress due to \Vtwo\ where unified formulation and exact solution for all effectively
2d backgrounds was given, earlier work in relation to Seiberg-Witten curve \ref\NN{N. Nekrasov, {\it
Seiberg-Witten Prepotential From Instanton Counting,}, hep-th/0206161,
Adv.Theor.Math.Phys. 7 (2004) 831-864.}, etc.; we also would like to note that M. Kontsevich have made
several proposals in this direction over the years \ref\Maxim{M. Kontsevich, 1993-2004, unpublished.}).
One interesting way to look at this is to find a direct link between
  the Kontsevich type representation \Kon\ and the KS functional
integral. The implications of all these  will be considered in \GStwo.
Let us only remark here that the natural setting  that emerges
is very much in the spirit of geometric considerations of
 \TT. Briefly - for $d$-dimensional CY manifold  $M$
 we  start with the  theory on $d+1$ dimensional Fano manifold
 $N$ together with the inclusion $M\subset  N$, such that the canonical
divisor of $N$ is
 proportional to the class of $M$ in $N$.
 The dynamics of the
 theory on $N$ is captured by  the effective theory on the  codimension
one
 sub-manifold - $M$. The codimension one sub-manifold
 may be considered as a complex analog of the boundary and one
expects the emergence of the phenomena familiar from AdS/CFT
correspondence (perhaps this can be compared to
\ref\GopV{R. Gopakumar and C. Vafa, {\it Topological Gravity as Large N Topological Gauge Theory,}
hep-th/9802016, Adv.Theor.Math.Phys. 2 (1998) 413-442.}).

\newsec{Appendix 1. Geometry of three-forms in six dimensions}

In this appendix we provide  brief description of the relevant
properties of three-forms in six dimensions following Hitchin \Hit
\Hitone.
Let $M$ be a six-dimensional Calabi-Yau  manifold. One of
the special properties of  three-forms in six dimensions is the
possibility to decompose any generic real three-form as a sum of two
(possibly complex conjugate) decomposable three-forms:
 \eqn\split{
 \Omega=\Omega_++\Omega_-=E^1\wedge E^2\wedge E^3+E^4\wedge E^5\wedge
E^6,}
where $E^i$'s are some one-forms. Generically $\Omega_-\wedge \Omega_+\neq
0$ and $E^i$'s produce the frame in
the complexified cotangent bundle $T^*_{\C}M$ to $M$. The
representation \split\ follows from the existence of
the open orbit of the group $GL(V)$ acting on  $\wedge^3 V^*$
where  $V$ a six-dimensional vector space.

One could explicitly  reconstruct the decomposable forms $\Omega_{\pm}$ as
follows.
Let ${\cal A}^{p}_M$ be the space of the real  $p$-forms
on $M$ and $Vect_M$ be the space of vector fields. Given a
three-form $\Omega$, consider the operator
$K_{\Omega}$:
  $$
K_{\Omega}:Vect_M\rightarrow {\cal A}^{5}_M\simeq Vect_M\otimes
{\cal A}^{6}_M,$$
defined as:  $$
v \rightarrow  (v\vdash \Omega) \wedge \Omega.
$$
For instance, given the decomposition \split\  with real $E^i$'s, the
action of $K_{\Omega}$
on the dual real frame $E^*_i$   is:  $$
 K_{\Omega}:E^*_i \rightarrow E^*_{i}, \,\,\, i=1,2,3,$$ $$
 K_{\Omega}:E^*_i \rightarrow -E^*_{i},\,\,\, i=4,5,6.$$
Now  the decomposable components $\Omega_{\pm}$  may be defined  as
follows. Let $K^*_{\Omega}$ be the group action of $K_{\Omega}$  on ${\cal
A}^p(M)$. Then: $$
2\Omega_+=
\Omega+ \l(\Omega)^{-3/2} K_{\Omega}^*\Omega,$$ $$ 2\Omega_-=
\Omega-\l(\Omega)^{-3/2}K_{\Omega}^*\Omega, $$
where:
\eqn\tr{
\l(\Omega)=\frac{1}{6} tr \, K_{\Omega}^2 \in
({\cal A}^6_M)^{\otimes 2}.}
Decomposition \split\ is non-degenerate (i.e $\Omega_+\wedge
\Omega_-$ has no zeros) if the form $\l(\Omega)$ has no zeros.
The sign of $\l(\Omega)$   defines whether
$\Omega_{\pm}$ are
real ($\l(\Omega) >0$) or complex conjugate to each other
$\Omega_-=\overline{\Omega}_+$ ($\l(\Omega) <0$). Denote the
corresponding subspaces in the space of real three-forms by $U_+$ and
$U_-$.
In the case $\Omega \in U_-$ the
operator $I$
 $$
 I_{\Omega}=(-\l(\Omega))^{-1/2}K_{\Omega},$$
defines the (pseudo)complex structure. The condition of the integrability
of
this complex structure may be described as
 $$ d\Omega=d\widehat{\Omega}=0,$$
where $\widehat{\Omega}\equiv \Omega_+-\Omega_-$. Note that
$\Omega+i\widehat{\Omega}$ is a holomorphic $(3,0)$-form without zeros
in this complex structure.
It turns out that the integrability condition may be formulated
as an equation for the critical points of the functional $\Phi(\Omega)$
written in terms of
the
closed three-form $\Omega$, $d\Omega=0$
\eqn\actone{\Phi(\Omega)=\int_M \sqrt{|\l(\Omega)|}.}
The variation of this
functional
is given by:
 $$ \delta
\Phi(\Omega)=-\int_M \widehat{\Omega} \wedge \delta \Omega. $$
The following relation holds  $$
\Omega_+\wedge \Omega_- =\frac{1}{2}\Omega\wedge \widehat{\Omega}= (\l
(\Omega))^{1/2}.$$
Let us restrict the space of three-forms $\Omega$ by the condition:
$d\Omega=0$,
and fix the class of $\Omega$ in $H^3(M,\R)$. Such three-forms
may be parameterized by two-form $\phi$ as:
$\Omega=x+d\phi$, where $x$ is some fixed closed three-form $[\Omega
-x]=0$ in $H^3(M,\R)$. The critical points of the functional
\actone\ under the
variation $\delta\Omega=d\delta \varphi$ are given by the
solutions of the equation $d\widehat{\Omega}=0$.
Given the cohomology class of the real closed three-form on $M$,
and using the critical point condition one could reconstruct unambiguously
 the holomorphic structure and the holomorphic
non-degenerate $(3,0)$-form $\Omega+i\widehat{\Omega}$ on $M$; thus
$M$ has a trivial canonical class. One can show \Hit\ that up
to
the action of the diffeomorphisms the critical point is isolated,
so it defines the map of the subspace of $H^3(M,\R)$ (such that the
corresponding critical value $\Omega$ is in $U_-$) into the extended
moduli space of complex structures $\widehat{{\cal M}}_M$.

\newsec{Appendix 2: On six dimensional field theory}

In this appendix we consider a  field
theory of two-forms that may be formulated due to the special
properties of six dimensional geometry. The exact quantum construction,
if exists, obviously  deserves the introduction of the additional degrees
of freedom. Thus our discussion will be rather  formal.

Consider the following formal path integral:
\eqn\Pol{
 Z=\int_{(x+d\phi)\in U_-} d\phi\,\int Dk \,\,e^{\int_M
\frac{1}{\sqrt{\frac{1}{6}tr k^2}}
(x+d\phi)k(x+d\phi)},}
 where
 $x$ is some fixed element of $H^3(M,\R)$,  $\phi$ is a
 two-form and $k \in End(T^*M)$  acts on arbitrary differential form
 as an element of the Lie
 algebra.
 The equations of motion for $k$ are algebraic and its
 solution is
  $k=\rho K_{\Omega}$ for $\Omega=(x+d\phi)$ with $\rho$ being an
  arbitrary non-zero function.  Substituting this  solution
  into the action in \Pol\ one finds that (in the classical
  approximation over $k$)  the  theory described by \Pol\ is equivalent
 to:  \eqn\Nam{
 Z=\int_{U_-} D\phi \,\,e^{\sqrt{-\l(x+d\phi)}},}
Note that \Nam\ does not depend on $\rho$.

This should be compared with well-known procedure in two
dimensions.  Mainly - start with Polyakov formulation of the string
moving in $d$ dimension: \eqn\Poltwod{
 Z=\int \left(\prod_{a=1}^p\,d\phi^a\right) \,Dg_{ij} \,\,e^{\int_{M}
   \sqrt{g}g^{ij}\sum_{a=1}^p
\pr_i \phi^a \pr_j\phi^a},}
In two dimensions the analog of $k$ can be explicitly
described in terms of the metric as:  $$
 k_i^j=|g|\epsilon_{ik}g^{k j}$$ One has  $tr\,k^2=2|g|$ (compare with
\tr).
Then the   action in 2d is given by  $$
 S=\int_{M} \frac{1}{\sqrt{\frac{1}{2} tr k^2}} \sum_{a=1}^p
 d\phi^a\,\wedge  (k
d\phi^a). $$
One can get rid of $k$ using  its  equations of motion, so result is a
Nambu-Goto
action: $$
 S=\int_{M} \sqrt{\det_{i,j=1,2}(\sum_{a=1}^p \pr_i \phi^a \pr_j
\phi^a)}.$$
Thus  we have a remarkable analogy between the case of the two-forms
in six dimensions and scalars in two dimension.

Note that the proper generalization of the metric in
two dimensions in this context is given by the
non-normalized operator of the complex structure $k$.

Let us finally remark that  higher
dimensional version of the  case of the several scalar fields $\phi^a$
$a=1,\cdots p$, in two dimensions is presumably
connected with the generalization of above approach
to
the $d$-dimensional manifolds with $N=dim\, H^{d/2,0}(M)>1$; the volume
$vol_M$ form
may be represented as $$
vol_M=\sum_{a=1}^N \Omega_a\wedge \overline{\Omega}_a$$
with $\Omega_a$ being the basis of $H^{d/2,0}(M)$.

We stress that in $6d$ case
the natural variable is not the metric but the volume form.
 Both in $d=2$  and $d=6$  case the actual variable is  the operator
 $k$  acting on $d/2$ dimensional forms. Locally in $d=2$ this operator is
  characterized by the normalization function. Due to the special
 properties of the metric in two dimensions the choice of the metric  is
also
 locally reduced to the choice of the conformal factor and thus is
 equivalent to the choice of the volume form.  On the
 contrary in six dimensions the parametrization of the metric is more
 involved and the proper generalization is in terms of the volume form
  and  the operator $k$.

\newsec{Appendix 3: Derivation of the Holomorphic anomaly equation}

In this appendix we give a  derivation of the Holomorphic
Anomaly (HA) equation  of \BCOV. Its interpretation as a  heat-kernel type
connection describing the change of the polarization in the quantization
of $H^3(M,\R)$ was proposed in \W\ (see also \AT\ for some additional
clarifications).

The main ingredient in \BCOV\ is the
fact that the moduli space of the (gauged) complex CY manifolds has
the structure of the (projective) special K\"{a}hler geometry
introduced first in the physical context by \dWvP\ (see \F\ for  more
rigorous presentation).
Below we derive HA equations in the case of an  abstract
(projective) special manifold  and then show how the specification
to the particular case of the moduli space at hand leads to the results
from
\BCOV. In abstract terms - starting with  the canonical flat connection
on the tangent bundle to the special K\"{a}hler manifold we construct
the flat connection on the associated  flat infinite dimensional
non-linear bundle of the Weyl algebras (quantization of
the formal completion  of the zero section of the
tangent bundle). The wave functions are given by the flat sections of
this connection (Holomorphic anomaly equation).  The
constructed  connection has interesting properties. For instance its
quasi-classical approximation defines new holomorphic structure on
the formal completion of zero section of the tangent bundle. This
new complex structure (as was shown by Kapranov in \Kap\ )   coincides
with the
complex structure on the formal completion of the natural
inclusion $M\rightarrow M \times \overline{M}$.

We start with the description of the local geometry  of the
(projective) special K\"{a}hler manifolds.  The   special  K\"{a}hler
manifold
is a  K\"{a}hler manifold with a  flat torsion-free symplectic connection
on the
tangent bundle with some compatibility condition between complex
and flat structures.

More explicitly let $I$, $g$ and $\omega$  be a
complex structure, a K\"{a}hler metric and corresponding K\"{a}hler form
on the
manifold $M$. Then $M$ admits a special K\"{a}hler structure if there is a
flat
torsion-free symplectic  connection ${\cal D}$ such that the following
compatibility condition holds:
 \eqn\comp{{\cal D}_{\mu}I_{\nu}^{\lambda} ={\cal
D}_{\nu}I_{\mu}^{\lambda}.}
The flat connection ${\cal D}$ defines the covering of $M$ by the flat
Darboux coordinate systems $(x^i,y_i)$:  $$ \omega =
\sum_{i} dx^i\wedge dy_i, $$ subjected to  the conditions:  $$
  {\cal D}\,dx^i ={\cal D}\,dy_j =0.
 $$

Let $\pr$ be a holomorphic part of the exterior derivative $d$.
From the compatibility condition \comp\ it follows that in the
decomposition:
 $$
 \pr =\sum dz^i\frac{\pr}{\pr x^i}-\sum dw_i\frac{\pr}{\pr y_i},
 $$
the one-forms $dz^i$ and $dw_i$ are holomorphic and locally exact.
The functions $z^i$ may be considered as  local coordinates and  one has
the following  expression for the corresponding vector fields:
 $$
 \frac{\pr}{\pr z^i} = \frac{1}{2}(\frac{\pr}{\pr x^i}-\tau
 _{ij}(z)\frac{\pr}{\pr y_i}), $$
where: \eqn\sym{
 \tau _{ij}  =  \frac{\pr w_i(z)}{\pr z^j}.}
The K\"{a}hler form $\omega$ has the type $(1,1)$ and
 therefore $\tau_{ij}$   is symmetric: $$ 0=\omega(\frac{\pr}{\pr
z^i},\frac{\pr}{\pr z^j})=\tau_{ij}-\tau_{ji}.$$
 Together with \sym\ it means that locally
there exists a holomorphic function ${\cal F}(z)$ with the
property: $$ \tau _{ij}(z)=\frac{\FF}{\pr z^{i} \pr z^{j}}. $$

The expression for the
K\"ahler form is easily obtained by evaluating $\omega(\frac{\pr}{\pr
z^i},\frac{\pr}{\pr \overline{z}^{i}})$. In terms of
holomorphic function ${\cal F}$ K\"ahler potential
$K$, K\"{a}hler form $\omega$ and K\"{a}hler
metric $g$ are given by: $$ K=\sum_i(z^i
\frac{\pr\overline{{\cal F}}}{\pr
\bar{z}^{\bar{i}}}M_i^{\bar{i}}-\overline{z}^{\bar{i}}
\frac{\pr{\cal F}}{\pr z^i}M^i_{\bar{i}}),$$
$$ \omega =\pr\apr
K=\frac{\sqrt{-1}}{2} \sum _{i,\bar{j}}Im \frac{\pr^2{\cal F}}{\pr
z^i
\pr z^j}dz^i\wedge dz^{\bar{j}}M^j_{\bar{j}}, $$
\eqn\metr{ g = \frac{1}{2} \sum _{i,\bar{j}}Im
 \frac{\pr^2{\cal F}_0}{\pr z^i \pr z^j} dz^i\otimes
dz^{\bar{j}}M^j_{\bar{j}}.}
Here we use the spurious matrix
$M^i_{\bar{j}}=\delta^i_{\bar{j}}$,
  $M_i^{\bar{j}}=\delta_i^{\bar{j}}$ to make the contraction of the
  indexes more natural.

In the case of the special geometry the complex structure is not
necessary covariantely constant with respect  to the
flat connection (otherwise all special K\"{a}hler
 manifolds would be flat). This property may be characterized by
 introducing   the following
 holomorphic symmetric three-tensor
 \eqn\struct{ C_{ijk}\equiv C(\frac{\pr}{\pr z^i},\frac{\pr}{\pr
z^j},\frac{\pr}{\pr
     z^k})=
-4\omega(\frac{\pr}{\pr z^i}, {\cal  D}_{\frac{\pr}{\pr
z^j}}(\frac{\pr}{\pr z^k}))
 =\frac{\FFF}{\pr z^{i}\pr z^{j}\pr z^{k}}.}
Then the curvature $R=D^2$ of the Levi-Civita connection $D=d+\Gamma$
for the metric $g$ may be written as:   \eqn\con{ \Gamma^{j}_{ik}
=-\aGG^{j
\bar{j}}\pr_i
  \GG_{k \bar{j}}=C_{i k l}\aGG^{\bar{l} j} M^l_{\bar{l}}}
\eqn\conone{R^k_{i\overline{j}l}=-\pr_{\overline{j}}\Gamma^k_{il}=
 C_{i l p}\aGG^{p \bar{q}} \bar{C}_{\bar{q} \bar {j} \bar{m}}\aGG^{k
\bar{m}}.}

The simplest case  when the structure of the special K\"{a}hler
geometry naturally arises is the Lagrangian sub-manifolds of the
flat holomorphic symplectic manifolds \Hittwo. Let
$V=T^*\C^n=\C^{2n}$  be the complex symplectic vector  space
supplied with the  holomorphic symplectic structure and
(indefinite) metric $G$:
\eqn\hypeone{\Omega =\sum_{i=1}^n dz^i\wedge dw_i,}
\eqn\hypertwo{G=\sqrt{-1}\sum_{i,\bar{i}=1}^n( dz^i\otimes d\overline{w}_{\bar
{i}}M_i^{\bar{i}}- dw_i\otimes
 d\overline{z}^{\bar{i}}M^i_{\bar{i}}).}

The  holomorphic embedding  $\phi:M \rightarrow  V$
 such that the image of $M$ is a  Lagrangian submanifold
 provides $M$ with the induced structure of the pseudo-K\"ahler manifold.
 Conversely, any simply connected special K\"ahler
manifold
$(M,J,g,{\cal D})$  admits a K\"ahler Lagrangian embedding  $\phi:
M\rightarrow V$
inducing the data $(g,{\cal D})$.
Any holomorphic Lagrangian submanifold
may be
locally described as a graph of the locally exact holomorphic one-form
$d{\cal F}$. In the discussed case the  holomorphic function ${\cal F}$
coincides with the function
 entering the  local description of the abstract special K\"{a}hler
 manifold above.

In order to derive the HA  equation form \BCOV\  we need to
consider the  special K\"{a}hler manifold
supplied with the action of $\C^*$. This leads to the notion  of
the projective  special manifold (see \F) and  moduli space of the
gauged  CY manifolds posses exactly these properties.
Thus, let $M$ be a special manifold with free
$\C^*$-action leaving invariant the flat connection ${\cal D}$ and
 the symplectic structure. We denote by  $N$  the corresponding
 quotient manifold.
   Also let the special holomorphic coordinates on $M$  be
of the degree one and the function ${\cal F}$  - of degree two with
respect to  the action of $\C^*$.
 Then given  the special pseudo-K\"{a}hler metric
on $M$ of the  signature $(n,1)$, the induced metric on the
quotient space provides $N$ with the structure of the projective special
K\"{a}hler manifold.

Let $(z^0,z^1,\cdots z^n)$ be a  coordinate system on $M$;
 the $\C^*$-action is given by:
 $
(z^0,z^1,\cdots z^n)\rightarrow (\lambda z^0,\lambda z^1,\cdots
\lambda z^n).
 $
The non-homogeneous coordinates :
 $(y^0=z^0 ,y^i=z^i/z^0) $ provide the natural coordinates
$(y^1,\cdots ,y^n)$  on $N$ and   the   K\"{a}hler potential has  the
form:
 $$
K(z^0,z^1, \cdots z^n)=\mid y^0\mid ^2 k(y^1,\cdots ,y^n),
 $$
with some function $k(y)$. The components of the metric are given
by  $$
 h_{i\bar{j}} =\pr_i\apr_j k(y) \mid y^0\mid ^2, \,\,\,\,\,\,\,
 h_{0\bar{0}}= k(y), $$
$$
 h_{i\bar{0}}=\pr_i k(y)\bar{y}^0 ,\,\,\,\,\,\,\,
h_{0\bar{i}}=\pr_{\bar{i}} k(y)y^0. $$

Define  new frame in the tangent space  $TM$ as:
 \eqn\frame{
 v _0 = y^0\frac{\pr}{\pr y^0}, \,\,\,\,\,\,\,
 v _i = \frac{\pr}{\pr y^i}+\pr_i \log k(y) y^0\frac{\pr}{\pr y^0}.}
Then  the  K\"{a}hler metric in this frame $g=\pr
\apr K $ on $M$:
 $$
 g_{0\bar{0}}= k(y)|y^0|^2, \,\,\,\,  g_{i\bar{0}}=g_{0\bar{i}}=0,  $$ $$
 g_{i\bar{j}} =\mid y^0\mid ^2 (\pr_i\apr_j  k(y)  -\frac{\pr _ik
 \apr_{\bar{j}}k}{k}),$$
 defines the special  K\"{a}hler metric on the quotient
space $N$:
 $$
 G_{ij}=\frac{g_{ij}}{g_{00}}=\pr\apr \log k(y^i),
 $$
and the metric on  $\C^*$-bundle ${\cal L}$:
  $$
   \| s(y)\|_{\cal L}^2=|s(y)|^2k(y).
  $$
The expression for the curvature of the metric $G$ may be easily obtained:
  $$
R_{i\bar{j}l\bar{m}}=
G_{i\bar{j}}G_{l\bar{m}}+G_{i\bar{m}}G_{l\bar{j}}-C_{ilp}\overline{C}_{\bar{j}\bar
{m}\bar{p}}G^{p\bar{p}}. $$

Now let us remind the notion of the special coordinates  on
K\"{a}hler manifold \BCOV\ (for more rigorous treatment see \Kap).
Fix  a point $(z_*^i,\bar{z}_*^{\bar{i}})$ in the   K\"{a}hler manifold
$M$ and
let   $\omega=\pr \apr K(z,\bar{z})$   be  the K\"ahler form. Some
components of the curvature of the Levi-Civita connection in the
K\"ahler geometry are identically zero:
  $$ [D_i,D_j]=0. $$
This allows us for any point $(z_*,\bar{z}_*) \in M$ to chose such local
coordinates $(\eta,\bar{\eta})$,    based at this point ($z^i=z^i_*+\eta^i
+{\cal O}(\eta^2)$),  that all terms of the Taylor expansion of the
Christoffel symbols over $\eta^i$  are zero:
$$
\frac{\pr^n}{\pr \eta^{i_1}\cdots \pr \eta^{i_n}}
\Gamma_{ij}^k(z,\bar{z})|_{z=z_*}=0.$$
Explicitly  new  coordinates $\eta$ are given by: $$
\eta^j(z,z_*)=G^{j\bar{j}}(z_*\bar{z}_*)(\apr_{\bar{j}}K(z,\bar{z}_*)
-\apr_{\bar{j}}K(z_*,\bar{z}_*)), $$
where we imply the analytic continuation of the K\"{a}hler potential.
 Thus for the special K\"ahler metric \metr\ we have
 \eqn\spec{
 Im(\tau _{ij}(z_*))\eta
 ^j=(w_i(z)-w_i(z_*))-(\overline{\tau}_*)_{ij}(z^j-z_*^j),}
  where $(\tau_*)_{ij}=\tau_{ij}(z_*)$.
It is  useful to express the relation between  the
coordinates $z^i$ and  $\eta^i$ in the form of the following
differential   equations: $$
 \frac{\pr z^i(\eta;z_*)}{\pr \eta ^j}=\frac{\pr z^i(\eta;z_*)}{\pr z_*
 ^j}+\Gamma_{jl}^k(z_*,\bar{z}_*)\eta^l\frac{\pr z^i(\eta;z_*)}{\pr \eta
^k}.
$$

Each special K\"{a}hler manifold $M$ may  be
considered locally as a Lagrangian sub-manifold in the flat holomorphic
symplectic manifold $W$. In particular $W$ may be identified with the
complexified tangent space at some fixed point in $M$.
Let us chose some  holomorphic Darboux coordinates $(z^i,w_i)$ on $W$.
Now given the
special coordinates $\eta^i$ on $M$ based at the point $(z_*,\bar{z}_*)$
it is natural to introduce the additional coordinates  $\xi_i$ such that
$(\eta^i
;\xi_i)$ provides another set of Darboux
coordinates in $W$. Namely consider  the following  canonical
transformation
from the variables $(z^i,w_i)$ to the new variables $(\eta^i,\xi_i)$
\eqn\coord{
 \eta ^i=\aGGs^{i
\bar{j}}M^j_{\bar{j}}(w_j-w_j(z_*)-\overline{\tau}_{\bar{i}\bar{j}}(z_*)M_j^{\bar{j}}(z^j-z_*^j)),}
\eqn\coordone{ \xi_i=(w_i-w_i(z_*))-{\tau}_{ij}(z_*)(z^j-z_*^j).}
  The  coordinates $(\eta,\xi)$ are canonical variables and the Lagrangian
plain $(\xi_i=0)$ is
tangent to Lagrangian subspace $\MM$ at the point
$(z_*;\overline{z}_*)$. Inverse transformation is:
\eqn\eq{
w_i=w(z_*)_i+M_i^{\bar{i}}\overline{\tau}_{\bar{i}\bar{j}}(z_*)\aGGs^{k\bar{j}}\xi_k
+\tau_{ik}(z_*)\eta ^k,}
\eqn\eqone{z^i=z_*^i+\aGGs^{i \bar{k}}M^k_{\bar{k}}\xi_k+\eta ^i.}
 Note that the restriction on $M$ ( $w_i=w_i(z)\equiv
\frac{\pr {\cal F}}{\pr z^i}$) turns  $\eta^i $ into the special
coordinates \spec.
The generating function $S(z,\eta)$ of the canonical transformation
\coord\
$$ dS(z;\eta)=w_idz^i-\xi _id\eta ^i,$$
is given by:
\eqn\canon{S(z;\eta)=\frac{1}{2}\overline{\tau}_{\bar{i}\bar{j}}(z_*)
M_i^{\bar{i}}M_j^{\bar{j}}
(z^i-z_*^i)(z^j-z_*^j)-\frac{1}{2}
Im  (\tau_{ij}(z_*))\eta^i\eta^j+} $$+Im(\tau_{ij}(z_*))\eta^i(z^j-z_*^j)+
 +w_i(z_*)(z^i-z_*^i)-{\cal  F}(z_*).$$
The  change of the coordinates $(\eta,\xi)$ under the
infinitesimal change of the
base point $(z_*,\overline{z}_*)$ is given by the following
canonical transformation: \eqn\chan{
 \delta \eta^i=\Gamma_{kl}^i\eta^l\delta z_*^k-\delta z_*^i
 +\aGGs^{i \bar{j}}\bar{C}_{\bar{j} \bar{l} \bar{k}}
\aGGs^{p \bar{l}}\xi_p\delta
 \overline{z}_*^{\bar {k} },}
\eqn\chanone{\delta \xi_i=-\Gamma_{ki}^l\xi_l\delta z_*^k
 +C_{ijk}\eta^j\delta z_*^k.}

Now consider the quantization of the holomorphic symplectic manifold in
the polarization defined by the condition that the
wave function $\Psi(\eta)$ depends only on the coordinates $\eta
^i$. According to the standard quantization  rules one has
 $$
  \widehat{\eta^j} \rightarrow \eta^j,$$ $$
   \widehat{\xi_j}\rightarrow i\frac{\pr}{\pr \eta^j}.
   $$
The variation of the  wave  function under the change of the base point
$(z_*,\bar{z}_*)$  is given by the Bogolubov transformation corresponding
to \chan. Thus we have the following compatible system of equations
on the wave function $\Psi(\eta,z_*,\overline{z}_*)$:
 \eqn\HAEone{
  \frac{\pr \Psi}{\pr z_*^i}=
(\Gamma_{ik}^j \eta^k \frac{\pr }{\pr \eta^j}
 -\frac{\pr }{\pr
 \eta^i}+\frac{1}{2}C_{ijk}\eta^j \eta^k+
  +\frac{\pr}{\pr z^i}\log (\det Im\,\tau_*)) \Psi,}
\eqn\HAEtwo{ \frac{\pr \Psi}{\pr
 \overline{z}_*^i}=\bar{C}_{\bar{i} \bar{j} \bar{k}}\aGGs^{p \bar{j}}
\aGGs^{q
\bar{k}} \frac{\pr^2 \Psi}{\pr
 \eta^p \pr \eta^q}.}
Before we start discussing the connection of \HAEone,\HAEtwo\ with
Holomorphic anomaly equation in \BCOV\ let us make some remarks.

The quasiclassical limit of the equation \HAEtwo\ defines the new
complex structure on the infinite-dimensional bundle over the base
$M$. It appears that this complex structure is induced from
the natural complex structure on ${\cal O}(M\times \overline{M})$ in the
vicinity of the diagonal $M\rightarrow M\times \overline{M}$.

Let us note that the general solution of \HAEone, \HAEtwo\
 may be  constructed stating with
an arbitrary holomorphic function $\psi(z)$ on $M$. It can  be checked
by the direct calculation that the wave function given by the
following integral transformation:
 \eqn\wave{
 \Psi(\eta,z_*,\overline{z}_*)=
\int d^n z [det(\frac{\pr^2 S}{\pr z^j \pr \eta ^k})]^{\frac{1}{2}}
e^{\frac{i}{\hbar}S(z,\eta)}
 \psi(z),}
satisfies the equation \HAEone,\HAEtwo\ for an arbitrary holomorphic
function $\psi(z)$. This expression is obviously correct quasi-classically
and due to
the linearity of the canonical transformation \chan,\chanone\ gives
the exact unitary transformation of quantum wave
function (the determinant term in the measure of integration
correctly  takes into account that wave
functions are naturally half-densities on Lagrangian sub-manifolds).

Suppose  the holomorphic  function $\psi(z)$ has the following
quasi-classical expansion:
 $$
\psi(z)=\exp {\frac{1}{\hbar}{\cal F}_0(z)+\sum_k \hbar^k{\cal
F}_k(z)}, $$ with some holomorphic functions ${\cal F}_k(z)$.
It easy to show that if ${\cal F}_0$ coincides with the function
${\cal F}$
entering the description of the special geometry - the wave function:
$\Psi(\eta,z_*,\bar{z}_*)$ has the following expansion:
 \eqn\cond{
 {\cal F}_0(\eta,z_*,\overline
 {z}_*)=\frac{1}{6}\frac{\FFFs}{\pr z^{i}\pr z^{j}\pr
z^{k}}\eta^i\eta^j\eta^k+\cdots}
This should be compared to the properties of the classical KS action
functional \BCOV\ (see also \DVV).

In order to deduce the  Holomorphic anomaly equations of \BCOV\ we
should specialize \HAEone, \HAEtwo\ to the case of  the projective special
K\"{a}hler manifold.  Note that $\eta^i$ may be considered as linear
coordinates on the fiber of the tangent bundle $TM$ at
$(z_*,\bar{z}_*)$ and thus define the  basis
$E^i$ in the tangent space $TM$ at the point $(z_*,\bar{z}_*)$.
By using the nonhomogeneous coordinates $(y^0,\cdots
y^n)$ one can construct another bases $e^i$:
 $$
 E^i=y^0 e^i+y^i e^0, \,\,\,\,\,\, E^0= e^0, $$
such that the dual basis is given by:
$$
 E_i=\frac{1}{y^0} e_i, \,\,\,\,\,\,  E_0=e_0-\sum_k \frac{y^k}{y^0}e_k.$$
 Finally we define the third basis $(f^0,\cdots ,f^n)$:
 $$
e^i= f^i, \,\,\,\,\,  e^0= f^0+\sum _i f^i y^0\pr_i\log k(y), $$ $$
f_0= e_0, \,\,\,\,\,\, f_i=e_i+y^0\pr_i\log k(y) e_0. $$
We would like to rewrite  the equations \HAEone,\HAEtwo\ in the
coordinates $(y^0,\cdots ,y^n)$ on the base and in the linear
coordinates $\eta^i_1$ in the fiber, associated with the frame
 $(f^0,\cdots ,f^n)$.
 For the
 symmetric three-tensor $C$ and Christoffel symbols simple calculation
gives: $$
 C(y)_{ijk}=\frac{1}{(y^0)^3}C(z)_{ijk}, $$ $$
 C(y)_{0jk}=C(y)_{00k}=C(y)_{000}=0,$$
 $$
  \Gamma ^k_{ij}(g)=\Gamma ^k_{ij}(G)+\pr_i\log
k(y)\delta^k_j,\,\,\,\,\,\,\,
  \Gamma ^0_{i0}(g)=\pr_i\log k(y),$$
 $$
  \Gamma ^0_{ij}(g)=\Gamma ^k_{i0}(g)=0. $$
Finally we have the following system of equations on the wave
  function $\Psi(\eta_1,y_*,\overline{y}_*)$:
  \eqn\BCOVone{ \frac{\pr  \Psi}{\pr y_*^i}
=(\Gamma_{il}^j\eta^l\frac{\pr }{\pr \eta_1^j}
  -\frac{\pr }{\pr  \eta_1^i}+
\pr_i\log k(y)
(y_*^0\frac{\pr}{{\pr \eta_1}^0}-\eta_1^0\frac{\pr}{\pr
  \eta_1^0}-\eta_1^i\frac{\pr}{\pr \eta_1^i})
 +}$$
  +\frac{1}{2}(y_*^0)^2C(y)_{ijk}\eta_1^j \eta_1^k  +
   \frac{\pr}{\pr y_*^i}\log (\det Im \,\,\tau_*)  \Psi, $$
$$ y_*^0\frac{\pr \Psi }{\pr
   y_*^0}=(y_*^0\frac{\pr}{\pr \eta_1^0}-\eta_1^0\frac{\pr}
{\pr \eta_1^0}-\eta_1^i\frac{\pr}{\pr \eta_1^i} )\Psi,
    $$
and
 $$
 \frac{\pr \Psi}
{\pr  \overline{y}_*^{\bar{i}}}=\overline{C}_{\bar{i}\bar{j}\bar{k}}
\aGGs^{p \bar{j}} \aGGs^{q \bar{k}}
 \frac{\pr^2 \Psi }{\pr
 \eta_1^p \pr \eta_1^q}
 +\pr_j\apr_{\bar{i}} \log k(y)\eta_1^j
 (y^0\frac{\pr}{\pr \eta_1^0}-\eta_1^0\frac{\pr}{\pr \eta_1^0}-
 \eta_1^i\frac{\pr}{\pr \eta_1^i}) \Psi. $$
These should be compared with  the Holomorphic anomaly  equations \BCOV.

In the rest of the appendix we briefly discuss the special K\"{a}hler
geometry  of the moduli space of the gauged CY manifolds.
Let $M$ be a CY threefold and $H^3(M,\C)$ be the complexified
middle  cohomology of $M$. Fix the symplectic basis $\{\g^i\}$ in
$H_3(M,\Z)$:$$
 <\g^i_+,\g_j^->=\delta^{ij},\,\,\, <\g_i^-,\g_j^->=0,\,\,\,\,
<\g^i_+,\g^j_+>=0.$$
The periods of a three-form $\omega$  \eqn\periods{
 z^i=\int_{\g^i_+}\omega,\,\,\,\,  w_i=\int_{\g_i^-}\omega,}
provide  the natural coordinates  in $H^3(M,\C)$.

There is a natural hyperk\"{a}hler structure on $H^3(M,\C)$
defined by the holomorphic two-form:
\eqn\CYhsymp{ \omega^{2,0}(\delta_1 \Omega,\delta_2 \Omega)=\int_M \delta_1
 \Omega \wedge \delta_2 \Omega, }
and the K\"ahler form:
\eqn\CYkal{\omega^{1,1}(\delta_1 \Omega,\delta_2 \Omega)=\int_M \delta_1
 \Omega \wedge \delta_2 \overline{\Omega}.}
In terms of the coordinates $(z^i,w_i)$ they are given by
\hypeone,\hypertwo.

Let $\widehat{{\cal M}}_M$ be the  extended moduli space of the
gauged complex structures on CY manifold $M$. The period map defines the
 embedding  $\widehat{{\cal M}}_M\rightarrow   H^3(M,\C)$
such that the image of the point corresponding to CY manifold $M$ supplied
with the holomorphic three-form
 $\Omega$ has coordinates $ (\int_{\gamma^1_{\pm}} \Omega, \cdots
,\int_{\gamma^n_{\pm}} \Omega)$. It may be shown that the image of
 $\widehat{{\cal M}}_M$ is actually a  Lagrangian sub-manifold in
$H^3(M,\C)$ with respect to the holomorphic symplectic structure
\CYhsymp\ and thus in the coordinates  \periods\  may be described
in terms of some holomorphic function ${\cal F}(z)$ as follows:
$$
 z^i=\int_{\g_i^+}\Omega, \,\,\,\,\,\,\,
 w_i=\int_{\g_i^-}\Omega=\frac{\pr {\cal F}(z)}{\pr z^i}.$$
Given a local coordinates $x^a$ on the moduli space
$\widehat{\cal M} $ the set of the
 three-forms $ \Omega,\, \nabla_a \Omega,\,  \overline{\nabla_a \Omega},\,
 \overline{\Omega}$ provide a  basis in the complexified tangent
space to $\widehat{\cal M}$ (as in Section 1 we identify
 $\Omega^{-p,q}$ and $\Omega^{3-p,q}$ with the help of the holomorphic
 three-form $\Omega$).  Taking into account that the complexified
 tangent  space to $\widehat{\cal M}_M$ may be identified with
 $H^3(M,\C)$ we get a family of the bases in $H^3(M,\C)$ parameterized
 by  $\widehat{\cal M}_M$.

Now let us define the coordinates $(\eta,\xi)$  on the tangent space
to the point  $(z_*,\bar{z}_*)\in \widehat{\cal M}_M$ as:  $$
 G_{a\overline{a}}\eta^a= \int_M (\omega(z)-\Omega(z_*)) \wedge
\overline{\nabla_a
   \Omega},$$ $$
 \xi_a= \int_M (\omega(z)-\Omega(z_*)) \wedge \nabla_a \Omega, $$
where $G$ is a K\"{a}hler metric associated with the K\"{a}hler form
\CYkal.
It is easy to show that these coordinates are the special case of the
coordinates \coord,\coordone\ introduced above and the  KS
action on the classical solutions takes the form:
\eqn\claction{ S_{cl}=\int_M \Omega(x|z_*,\overline{z}_*)\wedge
 \Omega(0|z_*,\overline{z}_*)=\sum_i
 (z^iw_i(z_*)-w_i(z)z_*^i).}

It is useful to specialize this representation to the limiting  case
of the choice of  $(z_*,\bar{z}_*)$ at the point
of the maximal degeneration of the complex structure
$(z_{\infty},\bar{z}_{\infty})$ (see e.g. \DM). In this case
we have $\lim_{z_*\rightarrow z_{\infty}} \Omega =[\g_0]^{\vee}$
 where $[\g_0]$ is the three-form dual to the invariant cycle $\g_0$
 and the action takes the simple form:
\eqn\maxdeg{ S_{cl}=\int_M \Omega(x|z_*,\overline{z}_*)\wedge
 [\g_0]^{\vee}=\int_{\g_0} \Omega(x|z_*,\overline{z}_*).}
\listrefs
\end